\documentclass[twocolumn,amsmath,amssymb]{revtex4}
\usepackage{graphicx}
\usepackage{dcolumn}
\usepackage{bm}

\begin{document}
\title{Two-Way Quantum Number Distribution Based on Entanglement and Bell-State
Measurements}
\author{Sung Soon Jang}
\author{Hai-Woong Lee}
\affiliation {Department of Physics, KAIST, Korea}
\date{\today}

\begin{abstract}
A scheme is proposed by which two parties, Alice and Bob, can
securely exchange real numbers. The scheme requires Alice and Bob
to share entanglement and both to perform Bell-state measurements.
With a qubit system two real numbers can each be sent by Alice and
Bob, resulting in four real numbers shared by them. The number of
real numbers that can be shared increases if higher-dimensional
systems are utilized. The number of significant figures of each
shared real number depends upon the number of Bell-state
measurements that Alice and Bob perform. The security of the
scheme against individual eavesdropping attacks is analyzed and
the effects of channel losses and errors discussed.
\end{abstract}

\maketitle

Since the introduction by Bennett and Brassard in 1984 of the
first complete protocol(BB84) for quantum key
distribution\cite{BB84}, many proposals for its variations,
improvements and modifications have appeared and their
experimental implementations have been developed\cite{QC}. The
security of the BB84 protocol and its variations rely upon the
quantum-mechanical principle that information gain in an attempt
to distinguish between two non-orthogonal quantum states
introduces a detectable disturbance in the state of the system.
The sender, Alice, sends the signal to the receiver, Bob, in
states chosen randomly from two conjugate bases. The eavesdropper,
Eve, cannot guess the basis right every time, and her attempt to
measure the signal in the wrong basis inevitably introduces an
error to the key transmission.

An interesting alternative to the BB84 protocol is the scheme
based on entangled pairs first proposed by Ekert in
1991(E91)\cite{E91}. In its original version with spin
(polarization)-entangled particles, Alice and Bob perform spin
measurements along one of three directions. The measurement
direction is chosen randomly and independently of each other.
Measurement outcomes obtained when they measure along the same
direction can be used for key generation, while those obtained
when they measure along different directions are used to test
Bell's inequality. The security of the E91 protocol depends upon
the fact that eavesdropping reduces the degree of correlation
between the two members of the entangled pair and that this
reduction manifests itself as a reduction in the degree of
violation of Bell's inequality.

In this work we propose a scheme which allows two parties, Alice
and Bob, to simultaneously and securely exchange real numbers. As
in the E91 protocol, the scheme requires Alice and Bob to share
entanglement. Instead of performing measurements along randomly
chosen directions, however, Alice and Bob are required to perform
Bell-state measurements. The security of the scheme relies upon
the fact that eavesdropping changes the outcome of the Bell-state
measurements. This change in the outcome of the Bell-state
measurements originates from the reduction in the degree of
correlation between the two members of the entangled pair caused
by eavesdropping. In this respect, the proposed scheme may be
considered as a variation of the E91 protocol. The scheme,
however, involves no random choice of bases or directions. With a
qubit system, Alice and Bob each can send two real numbers to each
other, resulting in four real numbers shared by them. The number
of significant figures of each shared real number is determined by
the number of Bell-state measurements that Alice and Bob perform.
The protocol may thus be considered as a scheme to allow Alice and
Bob to securely share integers (or collection of digits) whose
length is determined by the number of Bell-state measurements they
perform. The digits they share can be used for key generation for
cryptographic purposes.

Let us suppose that Alice has an EPR(Einstein-Podolsky-Rosen)
source that emits a large number N($\gg$1) of entangled pairs one
by one at a regular time interval, each pair in the same Bell
state. The Bell state can be any of the four Bell states
\begin{subequations}
\begin{eqnarray}
 \left| {\Phi _{00} } \right\rangle_{AB}  = \left| {\Phi ^ +  } \right\rangle_{AB}  =
 \frac{1}{{\sqrt 2 }}\left( {\left| 0 \right\rangle_A \left| 0 \right\rangle_B  + \left| 1 \right\rangle_A \left| 1 \right\rangle_B } \right) \\
 \left| {\Phi _{01} } \right\rangle_{AB}  = \left| {\Phi ^ -  } \right\rangle_{AB}  =
 \frac{1}{{\sqrt 2 }}\left( {\left| 0 \right\rangle_A \left| 0 \right\rangle_B  - \left| 1 \right\rangle_A \left| 1 \right\rangle_B } \right) \\
 \left| {\Phi _{10} } \right\rangle_{AB}  = \left| {\Psi ^ +  } \right\rangle_{AB}  =
 \frac{1}{{\sqrt 2 }}\left( {\left| 0 \right\rangle_A \left| 1 \right\rangle_B  + \left| 1 \right\rangle_A \left| 0 \right\rangle_B } \right) \\
 \left| {\Phi _{11} } \right\rangle_{AB}  = \left| {\Psi ^ -  } \right\rangle_{AB}  =
 \frac{1}{{\sqrt 2 }}\left( {\left| 0 \right\rangle_A \left| 1 \right\rangle_B  - \left| 1 \right\rangle_A \left| 0 \right\rangle_B } \right)
 \end{eqnarray}
\end{subequations}
but for the sake of the concreteness of argument, we take it as
$|\Phi_{00}\rangle_{AB}$. Alice keeps the qubit A of each pair and
sends the qubit B to Bob.

Alice has in her possession another set of $N$ qubits, which we
denote by the subscript $\alpha$, each of which she prepared in
the state $|\psi\rangle_\alpha = a |0\rangle_\alpha + b
|1\rangle_\alpha$ ($|a|^2 + |b|^2 = 1$). Since Alice prepared the
qubits $\alpha$ in this state, she and only she knows what $a$ and
$b$ are, and she keeps them to herself. Alice performs a series of
N Bell-state measurements on each pair of qubits $\alpha$ and A.
On the other side, Bob has in his possession yet another set of
$N$ qubits, which we denote by the subscript $\beta$, each of
which he prepared in the state $|\psi\rangle_\beta = x
|0\rangle_\beta + y |1\rangle_\beta$ ($|x|^2 + |y|^2 = 1$). Since
Bob prepared the qubits $\beta$ in this state, he and only he
knows what $x$ and $y$ are, and he keeps them to himself. Bob also
performs a series of $N$ Bell-state measurements on each pair of
qubits $\beta$ and B. The experimental scheme is depicted
schematically in Fig.~1.

In order to find the probability $P_{ijkl}$ that Alice's
Bell-state measurement yields $|\Phi_{ij}\rangle_{\alpha A}$ and
Bob's Bell-state measurement yields $|\Phi_{kl}\rangle_{\beta B}$,
we expand the total wave function $|\psi\rangle_{\alpha \beta A B}
= |\psi\rangle_\alpha |\psi\rangle_\beta |\Phi_{00}\rangle_{AB}$
in terms of $|\Phi_{ij}\rangle_{\alpha A} |\Phi_{kl}\rangle_{\beta
B}$ as
\begin{equation}
\left| \psi  \right\rangle _{\alpha \beta AB} =
\sum\limits_{i,j,k,l = 0}^1 {\left| {\Phi _{ij} } \right\rangle
_{\alpha A} \left| {\Phi _{kl} } \right\rangle _{\beta B} V_{ijkl}
}.
\end{equation}
A straight forward algebra yields
\begin{subequations}
 \begin{equation}
 V_{0000}  = V_{0101}  = V_{1010}  = V_{1111}  = \frac{1}{{2\sqrt 2 }}\left( {xa + yb} \right)
 \end{equation}
 \begin{eqnarray}
 V_{0001}  = V_{0100}  = V_{1011}  = V_{1110}  = \frac{1}{{2\sqrt 2 }}\left( {xa - yb}
 \right)\\
 V_{0010}  =  - V_{0111}  = V_{1000}  =  - V_{1101}  = \frac{1}{{2\sqrt 2 }}\left( {xb + ya}
 \right)\\
 V_{0011}  =  - V_{0110}  = V_{1001}  =  - V_{1100}  = \frac{1}{{2\sqrt 2 }}\left( {xb - ya} \right)
 \end{eqnarray}
\end{subequations}

\begin{figure}
\includegraphics[width=7cm]{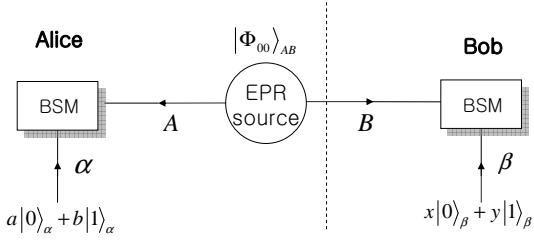} \caption{Experimental
Scheme. The EPR(Einstein-Podolsky-Rosen) source emits entangled
pairs in state $|\Phi_{00}\rangle_{AB}$. Alice performs Bell-state
measurements on the qubit pairs $\alpha$ and A, and Bob on the
qubit pairs $\beta$ and B. BSM stands for Bell-state measurement.}
\end{figure}

The probabilities $P_{ijkl}$'s are given by $P_{ijkl} =
|V_{ijkl}|^2$. Alice and Bob can determine these probabilities
experimentally from the result of their Bell-state measurements.
They only need to count the number $N_{ijkl}$ of occurrences for
the joint outcome $|\Phi_{ij}\rangle_{\alpha A}
|\Phi_{kl}\rangle_{\beta B}$. The experimentally determined
probabilities are then given by
\begin{equation}
P_{ijkl}^{exp }  = \frac{N_{ijkl}}{N}
\end{equation}

Suppose now, however, that Alice and Bob each announce publicly
her or his measurement result only when the outcome is $\Phi_{10}$
or $\Phi_{11}$. This is consistent with the realistic situation,
because only these two Bell states can be unambiguously
distinguished with linear optical means\cite{MWKZ96}. The
probabilities that can be determined experimentally are then only
$P^{exp}_{1010}, P^{exp}_{1011}, P^{exp}_{1110}$, and
$P^{exp}_{1111}$. These probabilities are given theoretically as
\begin{widetext}
\begin{subequations}
\begin{eqnarray}
 P_{1010}  = P_{1111}  = \frac{1}{8}\left| {xa + yb} \right|^2
 = \frac{1}{8}[\cos ^2 \theta _a \cos ^2 \theta _b
 + \sin ^2 \theta _a \sin ^2 \theta _b
 + 2\cos \theta _a \cos \theta _b \sin \theta _a \sin \theta_b \cos \left( {\phi _a  + \phi _b } \right) ] \\
 P_{1011}  = P_{1110}  = \frac{1}{8}\left| {xa - yb} \right|^2
 = \frac{1}{8}[\cos ^2 \theta _a \cos ^2 \theta _b
 + \sin ^2 \theta _a \sin ^2 \theta _b - 2\cos \theta _a \cos \theta_b \sin \theta _a \sin
\theta_b \cos \left( {\phi _a  + \phi _b } \right) ]
\end{eqnarray}
\end{subequations}
\end{widetext}
where we set
\begin{equation}
\begin{array}{l}
 a = \cos \theta _a ,\quad b = \sin \theta _a e^{i\phi _a }  \\
 x = \cos \theta _b ,\quad y = \sin \theta _b e^{i\phi _b }  \\
 \end{array}
\end{equation}
When the experimentally determined probabilities are substituted
for the corresponding theoretical probabilities, Eqs. (5)
constitute two equations that relate the four constants $\theta_a,
\phi_a, \theta_b$ and $\phi_b$. Since Alice knows $\theta_a$ and
$\phi_a$, she can use the two equations to solve for $\theta_b$
and $\phi_b$. Similarly, Bob knows $\theta_b$ and $\phi_b$, and
therefor he can use the two equations to solve for $\theta_a$ and
$\phi_a$. A third person, an eavesdropper, however, knows none of
the four constants, and there is no way for her to determine the
four unknown constants from the two equations. Thus, the method
described above, with Alice and Bob announcing her or his
measurement result only when the outcome is $\Phi_{10}$ or
$\Phi_{11}$, provides a means for Alice and Bob to securely share
four real numbers. Without loss of generality we take them as
$\cos\theta_a, \cos\phi_a, \cos\theta_b$ and $\cos\phi_b$, four
real numbers less than 1.

The number of real numbers that can be shared increases if Alice
and Bob use higher-dimensional systems. The generalized Bell
states for a d-dimensional system qudit) can be defined as
\cite{dBell}
\begin{equation}
|\Phi _{jl}\rangle_{AB}  = \frac{1}{\sqrt d
}\sum\limits_{q=0}^{d-1} {\omega ^{lq} } |q\rangle |q + j\rangle
\end{equation}
where $\omega=e^{i\frac{2\pi}{d}}$. As before, we assume that each
of the entangled pairs AB produced by the source is in
$|\Phi_{00}\rangle_{AB}$. Alice performs a series of Bell-state
measurements on each pair of the qudit A and another qudit
$\alpha$ she prepared in the state $\sum\limits_{i=0}^{d-1} a_i
|i\rangle_\alpha$, while Bob performs a series of Bell-state
measurements on each pair of the qudit B and another qudit $\beta$
he prepared in the state $\sum\limits_{i=0}^{d-1} x_i
|i\rangle_\beta$. As in the qubit case, the total wave function
$|\psi\rangle_{\alpha\beta A B}$ can be expanded in terms of the
Bell states $|\Phi_{ij}\rangle_{\alpha A} |\Phi_{kl}\rangle_{\beta
B}$, and the probability amplitudes $V_{ijkl}$'s can be expressed
in terms of $a_i$'s and $x_i$'s. we obtain
\begin{equation}
V_{ijkl} = \frac{1}{d\sqrt d} \:
\omega^{ij+kl}\sum\limits_{m=0}^{d-1} {\omega^{-(j+l)m}} a_{m-j}
\: x_{m-k}
\end{equation}
where all indices are evaluated modulo $d$. The probabilities
$P_{ijkl}$'s are then determined by $P_{ijkl}=|V_{ijkl}|^2$.

The constants $a_i$'s and $x_i$'s constitute (4d-4) unknowns to be
determined from experimentally determined probabilities
$P^{exp}_{ijkl}$'s. To Alice and Bob, however, there are only
(2d-2) unknowns. By agreeing to publicly announce the measurement
result only when the measurement outcome is among judiciously
chosen Bell states, Alice and Bob can limit the number of
probabilities that can be determined experimentally in such a way
that the number of equations that relate the experimentally
determined probabilities with the parameters $a_i$'s and $x_i$'s
is greater than or equal to (2d-2) but less than (4d-4). This way,
(4d-4) real numbers can be secretely shared between Alice and Bob.

\begin{figure}
\includegraphics[width=7cm]{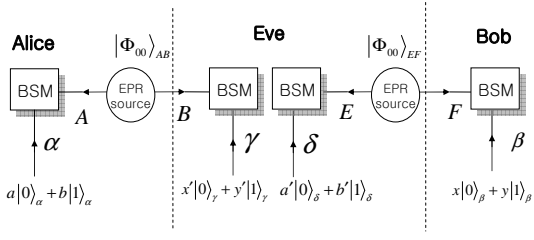} \caption{Eve's
intercept-resend attack}
\end{figure}

We now discuss the security of the scheme described above against
eavesdropping attacks. Perhaps the simplest attack that Eve can
attempt is the intercept-resend attack depicted in Figure~2. In
this attack Eve intercepts each qubit B being transmitted from
Alice to Bob and keeps it, while she generates her own entangled
pairs EF in the Bell state $|\Phi_{00}\rangle_{EF}$, keeps the
qubits E herself and sends the other qubits F to Bob. In addition
to qubits B and E, Eve has two sets of qubits $\gamma$ and
$\delta$, which she prepares in states, say, $x'|0\rangle_\gamma +
y'|0\rangle_\gamma$ and $a'|0\rangle_\delta + b'|1\rangle_\delta$,
respectively. Eve performs her own Bell-state measurements on the
pairs $\gamma$ and B and separately on the pairs $\delta$ and E.
By looking at correlations of the outcomes of her $\gamma$-B
measurement and Alice's $\alpha$-A measurement, Eve can determine
$a$ and $b$, i.e., $\cos \theta_a$ and $\cos \phi_a$. Similarly,
from correlations of the outcomes of her $\delta$-E measurement
and Bob's $\beta$-F measurement, she can determine $x$ and $y$,
i.e., $\cos\theta_b$ and $\cos\phi_b$, too. On the other side,
Alice and Bob would have performed their Bell-state measurements
on the pairs $\alpha A$ and $\beta F$, respectively. Note,
however, that the qubits A and F are not entangled, and thus
Alice's Bell-state measurement is completely independent of Bob's
Bell-state measurement. As a result, all the probabilities
$P_{ijkl}$'s should be the same, i.e.,
\[
P_{ijkl}=\frac{1}{16},~~~~~~~i,j,k,l = 0\mbox{ or }1
\]
Alice and Bob can check if $P_{1010}$ or $P_{1111}$ is the same as
(or close to) $P_{1011}$ or $P_{1110}$. If they feel that the two
probabilities are too close to trust, they discard the data and
restart from the beginning. It is possible that the two
probabilities are the same (or close) not because of Eve's attack
but because Alice and Bob happen to choose $\phi_a$ and $\phi_b$
such that $\cos(\phi_a+\phi_b) \simeq 0$. This case will also have
to be discarded. If Eve attacks not all but only a part of the
qubits B, the two probabilities may not be sufficiently close to
be detected. In this case Alice and Bob must resort to digit
comparison to detect Eve's attack. Due to the attack, the real
numbers $\cos\theta_a$ and $\cos\phi_a$ ($\cos\theta_b$ and
$\cos\phi_b$) computed by Bob(Alice) from the experimentally
determined probabilities will deviate from the correct values that
Alice (Bob) initially assigned. By comparing a few digits (e.g., a
digit at the third decimal point of $cos\theta_a$) and checking if
they agree, Alice and Bob can check against Eve's attacks. \\
\begin{figure}
\includegraphics[width=7cm]{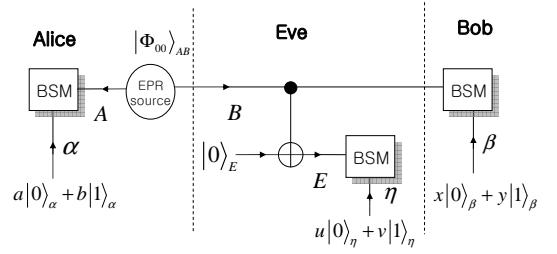} \caption{Eve's entangle-measure attack}
\end{figure}
\indent Another possible mode of attack is the
``entangle-measure'' attack depicted in Fig.~3. In this attack Eve
prepares a set of ancilla qubits E each in state $|0\rangle_E$,
entangle each of them with the qubit B by performing a CNOT
operation with the qubit B as the control bit and the qubit E as
the target bit, and performs a Bell-state measurement upon each
pair of the qubit E and another qubit $\eta$ a set of which she
prepares in state, say, $u|0\rangle_\eta + v|0\rangle_\eta$. In
this mode of attack, Eve's role is indistinguishable from Bob's
role, and she can obtain as much information as Bob can. In order
to find the effect of the entangle-measure attack upon the
probabilities $P_{ijkl}$'s, we expand the six-qubit wave function
$ |\psi\rangle_{\alpha \beta \eta ABE} = (a|0\rangle_\alpha   +
b|1\rangle_\alpha) (x|0 \rangle _\beta   + y|1\rangle_\beta)
(u|0\rangle_\eta + v|1\rangle_\eta) \frac{1}{{\sqrt 2 }}( {\left|
0 \right\rangle _A \left| 0 \right\rangle _B \left| 0
\right\rangle _E  + \left| 1 \right\rangle _A \left| 1
\right\rangle _B \left| 1 \right\rangle _E }) $ in terms of the
product of the Bell states as
\begin{equation}
\left| \psi  \right\rangle _{\alpha \beta \eta ABE} \, =
\sum\limits_{i,j,k,l,m,n = 0}^1 \left| {\Phi _{ij} } \right\rangle
_{\alpha A} \left| {\Phi _{kl} } \right\rangle _{\beta B} \left|
{\Phi _{mn} } \right\rangle _{\eta E} V_{ijklmn}
\end{equation}
and calculate the probabilities according to $ P_{ijkl}  =
\sum\limits_{m,n = 0}^1 {\left| {V_{ijklmn} } \right|^2 } $. A
straightforward algebra yields
\begin{subequations}
\begin{eqnarray}
\begin{array}{l}
 P_{0000}  = P_{0101}  = P_{1010}  = P_{1111}  =\\
 P_{0001}  = P_{0100}  = P_{1011}  = P_{1110}  = \frac{1}{8}\left( {\left| {xa} \right|^2  + \left| {yb} \right|^2 }
 \right)
\end{array} \\
\begin{array}{l}
 P_{0010}  = P_{0111}  = P_{1000}  = P_{1101}  =\\
 P_{0011}  = P_{0110}  = P_{1001}  = P_{1100}  = \frac{1}{8}\left( {\left| {xb} \right|^2  + \left| {ya} \right|^2 } \right)
\end{array}
\end{eqnarray}
\end{subequations}
In particular, the probabilities $P_{1010}, P_{1111}, P_{1011}$,
and $P_{1110}$ are all the same in this case and given by
\begin{equation}
\begin{array}{l}
P_{1010}  = P_{1111}  = P_{1011}  = P_{1110} = \frac{1}{8}\left(
{\left| {xa} \right|^2  + \left| {yb} \right|^2 } \right) \\
~~~~~~~= \frac{1}{8}\left( {\cos ^2 \theta _a \cos ^2 \theta _b +
\sin ^2 \theta _a \sin ^2 \theta _b } \right)
\end{array}
\end{equation}

The entangle-measure attack can thus be detected using the same
method employed to detect the intercept-resend attack. If Eve
entangles every qubit B with her ancilla qubit E, it can be
detected by checking if $P_{1010}$ or $P_{1111}$ is the same as
(or close to) $P_{1011}$ or $P_{1110}$. In general, however, Alice
and Bob should perform digit comparison to detect the attack,
because Eve can attack only a part of the qubits B.

Let us turn our attention to practical issues concerning the
proposed scheme. Suppose Alice and Bob want to securely share 4
real numbers less than 1 ($\cos\theta_a, \cos\phi_a, \cos\theta_b,
\cos\phi_b$) each accurate to D decimal points, or equivalently 4
integers each of length D, or equivalently 4D digits. How many
times do Alice and Bob each need to perform Bell-state
measurements? When a sufficiently large number N$\gg$1 of
Bell-state measurements are made, the number $N^{exp}_{ijkl}$ of
times the joint outcome $|\Phi_{ij}\rangle_{\alpha
A}|\Phi_{kl}\rangle_{\beta B}$ is counted lies within the range
defined as\cite{Reif}
\begin{widetext}
\begin{equation}
\begin{array}{l}
NP_{ijkl}  - \sqrt {2NP_{ijkl} \left( {1 - P_{ijkl} } \right)}
\lesssim N_{ijkl}^{exp} \lesssim NP_{ijkl}  + \sqrt {2NP_{ijkl}
\left( {1 - P_{ijkl} } \right)}
\end{array}
\end{equation}
\end{widetext}
where $P_{ijkl}$ is the exact theoretical probability given, for
example, for a qubit system by the absolute square of $V_{ijkl}$
given by Eqs. (5). Thus, the experimentally determined
probabilities $P^{exp}_{ijkl} = N^{exp}_{ijkl}/N$ are accurate to
$\sim \sqrt {\frac{{2NP_{ijkl} \left( {1 - P_{ijkl} }
\right)}}{N}}  \mathbin{\lower.3ex\hbox{$\buildrel<\over
{\smash{\scriptstyle\sim}\vphantom{_x}}$}} \frac{1}{{\sqrt N }} $
. Taking $N=10^n$, $P^{exp}_{ijkl}$'s are accurate to
$\frac{n}{2}$ decimal points. The real values $\cos\theta_a,
\cos\phi_a, \cos\theta_b$, and $\cos\phi_b$ that are determined
from these experimentally determined probabilities should also be
accurate to $D=\frac{n}{2}$ decimal points. We conclude therefore
that, for Alice and Bob to securely share 4D digits, they should
perform $\sim 10^{2D}$ Bell-state measurements each. The proposed
scheme has therefore a rather low efficiency of $\sim 4D10^{-2D}$.

The efficiency of the scheme can be enhanced by noting that the
efficiency decreases exponentially with D. Instead of trying to
obtain 4D digits in a single experiment consisting of $\sim
10^{2D}$ measurements, Alice and Bob can opt to divide it into
many independent experiments each with different values of
parameters $a$, $b$, $x$ and $y$. For example, consider the
situation where Alice and Bob want to share 400 digits. They can
achieve it by performing $\sim 10^{200}$ Bell-state measurements
in a single experiment and obtaining 4 real numbers accurate to
100 decimal points, i.e., 400 digits. Alternatively, they can
choose to shoot for only four digits in a single experiment by
performing $\sim 10^2$ Bell-state measurements and obtaining 4
real numbers accurate only to one decimal point. They can then
repeat the experiment 100 times, each time with different values
of $a$, $b$, $x$ and $y$ to obtain 400 real numbers each accurate
to one decimal point, i.e., 400 digits. Using this
``divide-repeat'' strategy, the number of Bell-state measurements
performed by Alice and Bob is reduced to $\sim 10^4$ and the
efficiency is enhanced to $\sim 400/{10^4} = 4 \times 10^{-2}$.
Even if some (two or three) of these four digits obtained from
each single experiment need to be used for checking against
eavesdropping attacks, the efficiency still remains to be $\sim
10^{-2}$. If Alice and Bob feel that they need more digits than
two or three from each single experiment to be used for the
security check, they can make $\sim 10^4$ Bell-state measurements
and obtain eight digits in a single experiment. They can then
repeat the experiment 50 times to obtain 400 digits altogether.
The efficiency in this case is $\sim 1.6\times 10^{-5}$. Another
way of increasing the efficiency is to use high-dimensional
systems. Since the number of real numbers that can be shared
increases with increased dimension, the efficiency also increases
by going to high-dimensional systems.

Up to now we have assumed an ideal situation where there are no
losses and no errors. In general, however, losses and errors are
unavoidable and their effects must be taken into account. Due to
losses, only $N\eta$ qubits out of $N$ qubits sent from Alice will
be detected by Bob, where $\eta$ is the probability that a single
photon sent form Alice is detected at Bob's detectors. If one
considers only the channel losses, it is given by $\eta =
10^{-(\alpha l + c)/10}$, where $\alpha$ is the absorption
coefficient, $l$ is the length of the channel(fiber) and $c$
accounts for a distance-independent loss in the channel. The
probabilities $P^{exp}_{ijkl}$'s should then be determined by
comparing the number $N^{exp}_{ijkl}$'s not to $N$ but to $N\eta$.
A more accurate determination of the probabilities can be obtained
if one lets Bob announce his measurement result every time he
receives a qubit B. He should announce whether the outcome of his
Bell-state measurement is $\Phi_{10}$ or $\Phi_{11}$ or
inconclusive (corresponding to the case where the outcome is
either $\Phi_{00}$ or $\Phi_{01}$ but he cannot distinguish
between the two). The number $N^{exp}_{ijkl}$ can then simply be
normalized to the number of times Bob has made his announcement.

Errors can occur during generation, transmission and detection of
qubits and can seriously limit the performance of our proposed
scheme. Under ideal errorless conditions, the number of digits
that Alice and Bob share can be increased simply by increasing the
number of qubits they prepare and the number of measurements they
perform. When errors are present, however, the error rate limits
the number of meaningful digits that Alice and Bob share through a
single experiment, and it may be meaningless to increase the
number of measurements to be made in a single experiment beyond a
certain level. For example, suppose the error rate is 5\%. The
accuracy of the probabilities $P^{exp}_{ijkl}$'s determined from
the experiment cannot be better than 5\%, which means that even
the digit at the second decimal point of the real numbers
$\cos\theta_a, \cos\phi_a,\cos\theta_b$ and $\cos\phi_b$
determined from these probabilities is not guaranteed to be
accurate. It is then best for Alice and Bob to shoot for four
digits, one digit for each real number, in a single experiment.
The number of measurements that can guarantee the accuracy of the
digit at the first decimal point is $\sim 10^2$ and it is in this
case meaningless to increase the number of measurements well
beyond $\sim 10^2$ in one experiment. When more digits are desired
to be shared, Alice and Bob need to repeat the process of $\sim
10^2$ measurements with different sets of parameters $a$, $b$, $x$
and $y$. Thus, the ``divide-repeat'' strategy is not only
desirable to enhance the efficiency but also required to make the
scheme work in the presence of errors.

We are now in a position to propose a protocol for quantum number
distribution which allows two parties, Alice and Bob, to share
securely a certain number of digits. Alice and Bob should proceed
as follows.
\begin{enumerate}
\item[(1)] Alice prepares $N$ entangled pairs AB, each in state
$|\Phi_{00}\rangle_{AB}$, keeps the qubit A and sends the qubit B
to Bob. Alice has another set of $N\eta$ qubits $\alpha$ and
divides them into $N\eta/100$ groups with each group consisting of
100 qubits. (We assume for simplicity that $N\eta$ is an integral
multiple of 100.) She prepares the qubits in the jth group (j=1,
2, ..., $N\eta/100$) in state ($a_j |0\rangle_\alpha + b_j
|1\rangle_\alpha$). Bob has a set of $N\eta$ qubits $\beta$ and
divides them into $N\eta/100$ groups with each group consisting of
100 qubits. He prepares the qubits in the jth group in state ($x_j
|0\rangle_\beta + y_j |1\rangle_\beta$). \item[(2)] Bob takes the
first 100 qubits B he receives and the 100 qubits in the first
group of the qubits $\beta$. He performs a Bell-state measurement
on each qubit pair $\beta B$ and announces publicly where the
outcome of each measurement is $\Phi_{10}$ or $\Phi_{11}$ or
inconclusive. Alice takes the 100 qubits A, entangled partners of
the first 100 qubits B that Bob received, and the 100 qubits in
the first group of qubits $\alpha$. She performs a Bell-state
measurement on each qubit pair $\alpha A$ and announces publicly
whether the outcome of each measurement is $\Phi_{10}$ or
$\Phi_{11}$. \item[(3)] Alice and Bob count the numbers
$N^{exp}_{1010}$, $N^{exp}_{1111}$, $N^{exp}_{1011}$, and
$N^{exp}_{1110}$ of joint occurrences of
$|\Phi_{10}\rangle_{\alpha A} |\Phi_{10}\rangle_{\beta B}$,
$|\Phi_{11}\rangle_{\alpha A} |\Phi_{11}\rangle_{\beta B}$,
$|\Phi_{10}\rangle_{\alpha A} |\Phi_{11}\rangle_{\beta B}$, and
$|\Phi_{11}\rangle_{\alpha A} |\Phi_{10}\rangle_{\beta B}$, and
determine the corresponding probabilities $P^{exp}_{ijkl} =
N^{exp}_{ijkl} / 100$. From the probabilities, they determine
$cos\theta_a, cos\phi_a,cos\theta_b$ and $cos\phi_b$, each to the
first decimal point. They now share 4 digits. \item[(4)] As a
check for the accuracy of the experiment, Alice and Bob check if
$P^{exp}_{1010}$ and $P^{exp}_{1111}$ agree at least to the first
decimal point. If not, they discard the data and restart. They do
the same checking for $P^{exp}_{1011}$ and $P^{exp}_{1110}$. As a
check against eavesdropping attacks, they check if
$P^{exp}_{1010}$ (or $P^{exp}_{1111}$) is sufficiently different
from $P^{exp}_{1011}$ (or $P^{exp}_{1110})$. If not, they discard
the data and restart. \\
As a further check against eavesdropping attacks, Alice and Bob
each take two of the four digits they share ( they could take one
or three digits depending upon the level of confidence) and
publicly compare and check if each of the two pairs agree. If the
agreement is found, then they each keep the remaining two digits
as the key. If not, they discard the data and restart. \item[(5)]
The steps (2)-(4) are repeated $\frac{N\eta}{100}$ times, each
time with a different set of 100 qubits each of A, B, $\alpha$ and
$\beta$. When all measurements are completed successfully, Alice
and Bob have collected between them $2\times \frac{N\eta}{100}$
real numbers each accurate to the first decimal point, i.e.,
$2\times \frac{N\eta}{100}$ digits. The $2\times
\frac{N\eta}{100}$ digits constitute the final key.
\end{enumerate}

We note that if the error rate is below 1\%, Alice and Bob can
shoot for 8 digits instead of 4 digits in a single experiment, by
dividing the qubits into groups of $10^4$ qubits instead of 100
qubits, and performing $\sim 10^4$ Bell-state measurements instead
of 100 measurements in a single experiment. Each single experiment
will then produce 4 real numbers accurate to the second decimal
point, i.e., eight digits. This way Alice and Bob have more qubits
available for digit comparison, but the efficiency will be lower.
\\
\indent In conclusion we have proposed a protocol based on
entanglement and Bell-state measurements that allows two parties
to exchange real numbers securely. As compared with the standard
quantum cryptographic protocols such as BB84, our proposed
protocol suffers from the low efficiency. With the help of the
``divide-repeat'' strategy, however, its efficiency can be
increased to $\sim 10^{-2}$. As the security of the proposed
protocol relies upon the fact that an act of eavesdropping changes
the outcome of the Bell-state measurements, the protocol requires
the process of digit comparison to protect against eavesdropping
attacks. The protocol, however, does not require random choice
between two conjugate bases as in BB84 nor the Bell's inequality
test as in E91. The proposed protocol appears to protect itself
well against eavesdropping attacks. It is secure, in particular,
against an individual attack in which Eve attacks every qubit
transmitted from Alice to Bob, because such an all-out attack
leaves its mark on the probabilities. If Eve attacks only a part
of the qubits, then Alice and Bob have to perform information
reconciliation, which consists of checking if some randomly
selected digits they share agree. Ironically, the low efficiency
of the protocol works to help this digit comparison process
effective. Because of the low efficiency, the information on
whether there were eavesdropping attacks is contained in a
relatively small number of digits produced by the protocol. The
agreement between just a small number of pairs of digits can thus
be considered as a strong
indication for the absence of eavesdropping attacks. \\
\indent We note that our proposed protocol provides a way of
two-way communication, allowing simultaneous mutual exchange of
information between Alice and Bob. Alice and Bob are
simultaneously both the sender and the receiver of information,
while in standard cryptographic protocols information usually
flows one way. \\
\indent On the practical side, a successful operation of the
proposed protocol requires generation, distribution, and detection
of entanglement at a single-photon level, a difficult but not an
impossible task. It requires, in particular, a large number of
Bell-state measurements to be performed. We emphasize, however,
that only two of the four Bell states are required to be
distinguished. The distinction of the two Bell states is possible
using only linear optical means and therefore can be accomplished
without too much difficulty with the present technology. \\
\indent Channel errors must be minimized for a successful
operation of the protocol. If our protocol is to work at all, the
error rate must be kept below $\sim 10\%$, because the error rate
of over $\sim 10\%$ will not guarantee the accuracy of the digit
even at the first decimal point.


\begin{thebibliography}{99}
  \renewcommand{\baselinestretch}{1}
  \selectfont
\bibitem{BB84}
  C.~Bennett,and G.~Brassard, in {\em Proceedings of IEEE International
  Conference on Computers, Systems, and Signal Processing, Bangalore,
  India} (IEEE, New York, 1984), p. 175
\bibitem{QC}
  For a review, see, for example, N. Gisin, G. Ribordy, W. Tittel,
  and H. Zbinden, Rev. Mod. Phys. \textbf{74}, 145 (2002).
\bibitem{E91}
  A. K. Ekert, Phys.Rev.Lett. \textbf{67}, 661 (1991).
\bibitem{MWKZ96}
  K. Mattle, H. Weinfurter, P. G. Kwiat, and A. Zeilinger, Phys.Rev.Lett. \textbf{76}, 4656 (1996)
\bibitem{dBell}
  G. Alber, A. Delgado, N. Gisin, and I. Jex, e-print
  quant-ph/0008022; S. J. van Enk, Phys.Rev.Lett. \textbf{91},
  017902(2003)
\bibitem{Reif}
  See, for example, F. Reif,{\em Fundamentals of Statistical and
  Thermal Physics} (McGraw-Hill, New York, 1965), Ch. 1.
\end{thebibliography}
\end{document}